# Metallic beta-phase silicon nanowires: structure and electronic properties

Pavel B. Sorokin[*, 1, 2, 3], Pavel V. Avramov[3], Viktor A. Demin[1], and Leonid A. Chernozatonskii[2]

[1]*Siberian Federal University, 79 Svobodny av., Krasnoyarsk, 660041 Russian Federation*
[2]*Emanuel Institute of Biochemical Physics, Russian Academy of Sciences, 4 Kosigina st., Moscow, 119334, Russian Federation*
[3]*Kirensky Institute of Physics, Russian Academy of Sciences, Akademgorodok, Krasnoyarsk, 660036 Russian Federation*
**\*** Corresponding author: PSorokin@iph.krasn.ru

Electronic band structure and energetic stability of two types of $\langle 110 \rangle$ and $\langle 001 \rangle$ oriented silicon nanowires in β-Sn phase with the surface terminated by hydrogen atoms were studied using density functional theory. It was found that β-Sn nanowires are metastable with zero band gap against to α-diamond nanowires. The relative energy of the studied wires tends to the energy of the bulk silicon crystal in β-Sn phase.

Silicon nanowires (SiNW) are the most perspective elements of nanotechnology. SiNW can be used in nanoelectronics as field-effect transistors, diodes, logic gates, and more. A huge number of theoretical [1]-[7] and experimental [8], [9] papers devoted to investigation of properties of SiNW were published in the last years.

The structure and properties of silicon nanowires and nanorods under high pressure is almost a virgin field. Only two papers [10], [11] devoted to this topic were published. In these works a study of the silicon nanowires (SiNW) in different high-pressured phases was reported. In particular, it was found that the SiNWs start to transform from initial α-diamond to β-Sn structure in the range of 7.5 to 10.5 GPa [10], [11]. Despite of low activity of investigations of high pressured nanowires a plenty of experimental [12]-[15] and theoretical [16]-[19] studies of the high-pressure phases of bulk silicon were carried out.

An existence of silicon nanowires in new phases opens a new promising field in the nanomaterial science due to their useful electronic properties. In particular, the silicon nanowires in β-Sn phase have larger bulk modulus than the bulk module of corresponding crystal [11].

The main goal of this paper is an *ab-initio* study of hydrogen-covered thin α-diamond SiNWs as well as β-Sn SiNWs previously discovered in experiment [10], [11]. We believe that SiNWs in β-Sn phase could display metallic properties like bulk β-Sn silicon. The assumption was confirmed by *ab initio* calculations given below.

The electronic structure calculations of a set of silicon nanowires were carried out using density functional theory (DFT) in the framework of local density approximation [20]-[22] with periodic boundary conditions using Vienna Ab-initio Simulation Program [23]-[25]. We used a planewave basis set, ultrasoft Vanderbilt pseudopotentials [26] and a plane-wave energy cutoff equal to 200 eV. To calculate the equilibrium atomic structures, the Brillouin zone was sampled according to the Monkhorst–Pack [27] scheme with a 0.1 Å$^{-1}$ k-points density grid. To avoid interactions with the wire periodic image, the neighboring SiNWs were separated by 15 Å in the tetragonal supercells. During atomic structure minimization, structural relaxation was performed until the change in total energy was less than 10$^{-3}$ eV/atom.

At the DFT level of theory the lattice parameter of the β-Sn phases of bulk silicon was predicted with an error of 1.5% ($a_{calc}$ = 4.76 Å  $a_{exp}$ = 4.69 Å, [15]; $(c/a)_{calc} = 0.550$, $(c/a)_{exp} = 0.550 \pm 0.002$ [15], 0.554 [13]). The energy difference between the α-diamond and β-Sn phases of silicon is 0.22 eV/atom, which is in close agreement with previous DFT calculations of 0.22 [16] and 0.21 eV/atom [17].

The atomic structures of silicon in α-diamond ($m\bar{3}m$ cubic unit cell) and β-Sn phases (4/mmm tetragonal unit cell) are presented in Fig. 1a. Different symmetries dictate the different notations of the crystallographic directions of the crystals. a and b crystallographic axes of the β-Sn unit cell are rotated on 45º relative to the corresponding axes of the α-diamond cell. Due to this and $a = b \neq c$ in the tetragonal unit cell, the equivalent $\langle 100 \rangle$, $\langle 010 \rangle$ and $\langle 001 \rangle$ directions in α-diamond phase correspond to equivalent $\langle 110 \rangle$, $\langle \bar{1}10 \rangle$ and non-equivalent $\langle 001 \rangle$ directions in β-Sn phase. We studied the $\langle 110 \rangle$ (Fig. 1b) and $\langle 001 \rangle$ (Fig. 1c) oriented SiNWs truncated from the bulk silicon in β-Sn phase and compared them with $\langle 100 \rangle$ wires of different effective sizes truncated from the bulk silicon in α-diamond phase (Fig. 1d). The size of studied SiNW was chosen according to Ref. [9] where the nanometer-size silicon wires were produced. For example, the perspective views of $\langle 110 \rangle$ and $\langle 001 \rangle$ SiNWs in β-Sn phase are presented in Fig. 1e.

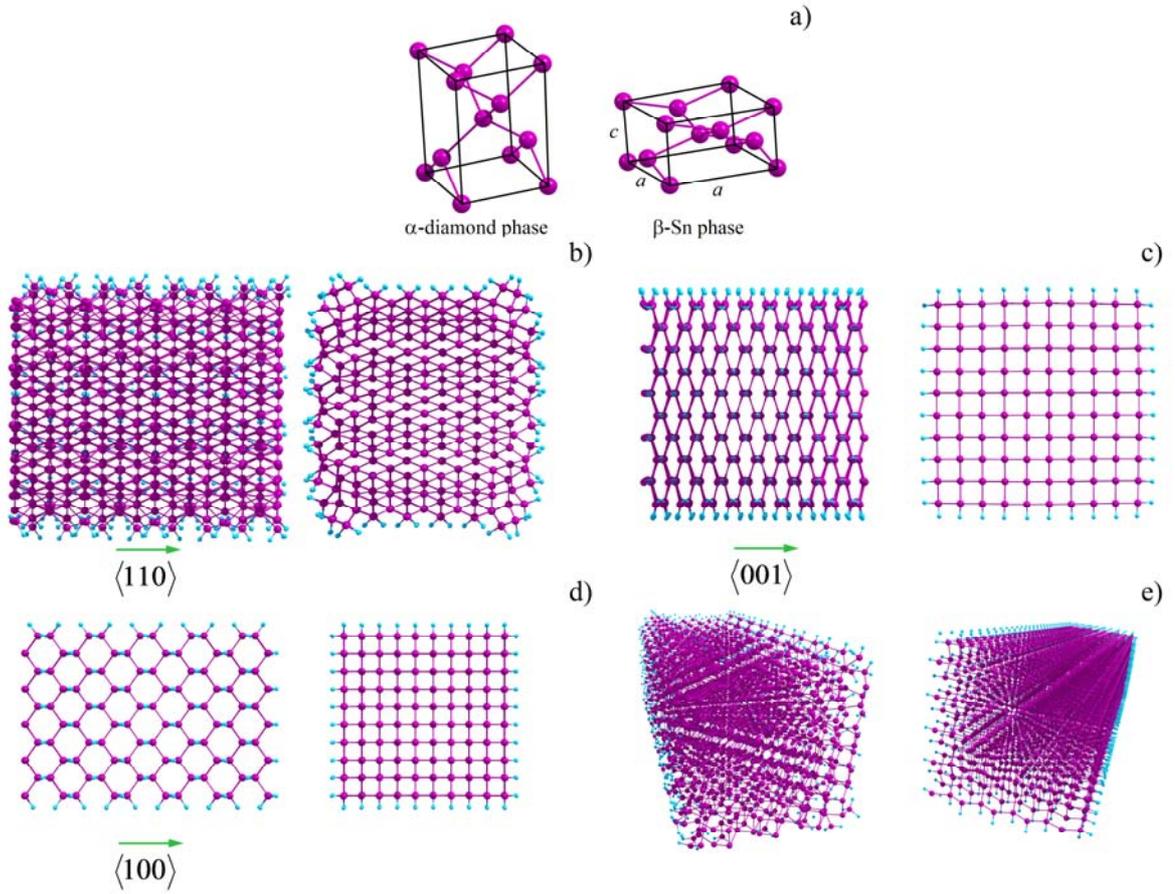

Fig. 1. a) The α-diamond and β-Sn silicon unit cells; Side and cross-section views of b) ⟨110⟩ SiNW in β-Sn phase, c) ⟨001⟩ SiNW in β-Sn phase and d) ⟨100⟩ SiNW in α-diamond phase; e) The perspective view of ⟨110⟩ and ⟨001⟩ SiNWs in β-Sn phase.

To study the electronic structure and energetic stability of SiNWs, the uniformly hydrogen-covered silicon nanowires were calculated. Hydrogen passivation prevents reconstruction of surface of silicon wires under the study and conserves the conducting properties of the materials. For example, unpassivated ⟨100⟩ oriented silicon nanowires in α-diamond phase reveal metallic properties, whereas hydrogen passivation restores the semiconducting band gap [7].

It was found that in comparison with the parent crystal, the deviation of the Si-Si bond lengths inside both ⟨001⟩ and ⟨110⟩ oriented β-Sn SiNWs is less than 1.5 % with wider distribution of the bond lengths on the wires surface. Orientation and the surface relaxation of the ⟨110⟩ β-Sn SiNWs determine wider bond length deviation on the surface (up to 4.4 %). The ⟨001⟩ oriented β-Sn wires display smaller surface relaxation with deviation of the bond lengths on the surface up to 2.7 %.

To determine the effective size of the nanowires under study, the area of perpendicular cross-section of the species [2], [3] was used. Some other methods, like the determination of a single effective diameter of a nanowire as the smallest diameter taken from a perpendicular cross section [4] or the smallest cylinder containing a nanowire [5], [6] are not unambiguous and cannot be applied to species with complex shapes and different facets. The cross-section of SiNW was taken as the polygonal region which was defined by the surface hydrogen atoms.

The 1D nature and high surface tension of the wires are expected to imply an increasing of the strain of the structures. The dependence of SiNWs relative energies upon the cross-section area is presented in Fig. 2. The nanowires relative energy is defined as energy per atom of a wire of given cross-section area relatively to the energy per atom of bulk silicon in α-diamond phase.

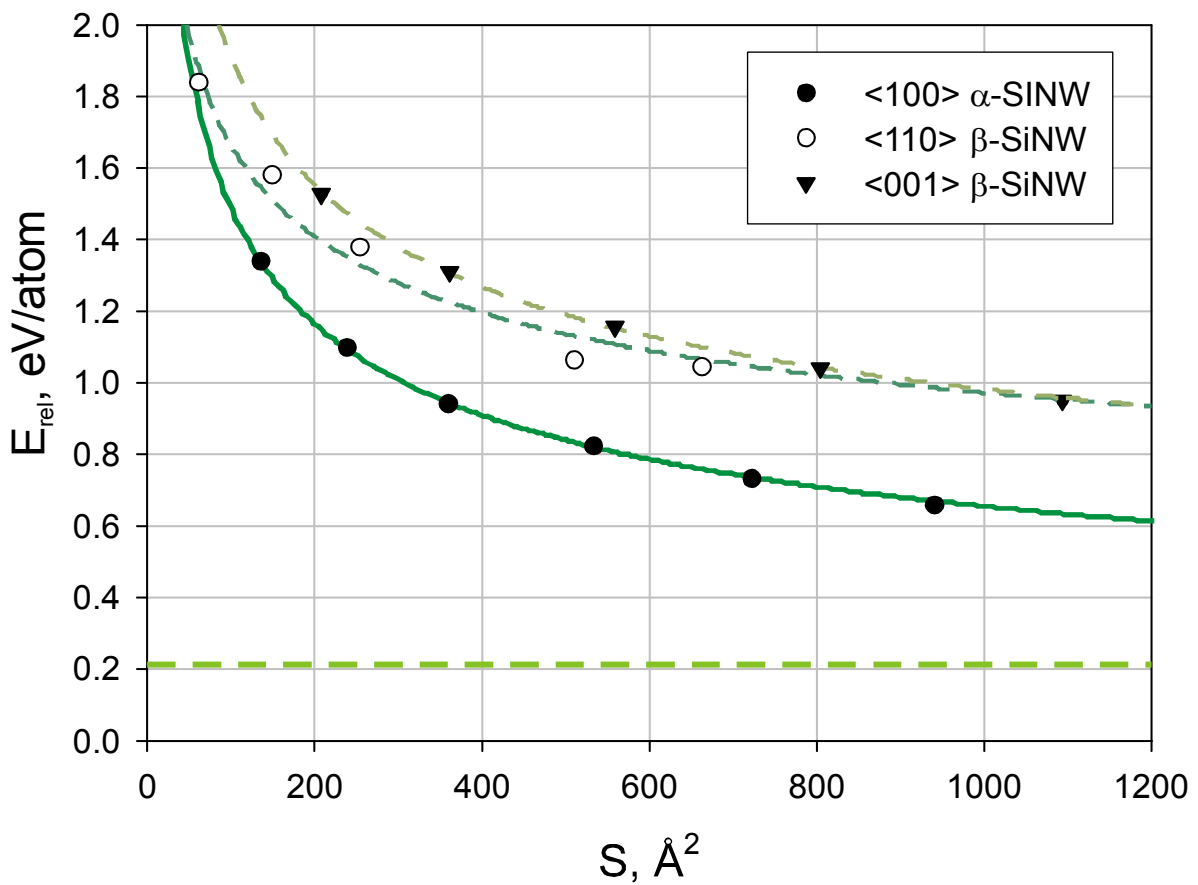

Fig. 2. The relative energy ($E_{rel}$) as a function of SiNW cross-section area. $E_{rel}$ of SiNWs in β-Sn phase of ⟨110⟩ orientation is marked by empty circles with dashed approximating curve. The SiNWs in β-Sn phase of ⟨001⟩ orientation is marked by filled triangles with dashed approximating curve. The relative energy of SiNWs in α-diamond phase is marked by filled circles with solid approximating curve. The relative energy of bulk silicon in β-Sn phase is marked by horizontal dashed line.

The dependence of nanowires relative energy upon the size can be approximated by the power law. The data points of the energy were fitted to $E_{rel} = A \cdot S^c + E_0$ where $E_0$ is the energy of the corresponding crystal, $S$ is the cross-section area, $A$ and $c$ are fitting parameters. The $E_0$ for SiNWs in α-diamond phase was defined as 0 and for SiNWs in β-Sn phase as 0.22 eV/atom. The $A$ and $c$ parameters calculated using $E_0$ values are presented in Table 1.

Table 1  $A$ and $c$ parameters of approximating curves of dependences of strain energies upon cross-section area of the studied SiNWs.

|   | SiNW in α-diamond phase oriented in $\langle 100 \rangle$ direction | SiNW in β-Sn phase oriented in | |
|---|---|---|---|
|   |   | $\langle 110 \rangle$ direction | $\langle 001 \rangle$ direction |
| $A$ | 7.859 | 5.373 | 8.439 |
| $c$ | -0.3603 | -0.285 | -0.3483 |

The SiNWs in α-diamond phase are energetically favorable the SiNWs in β-Sn phase with close cross-section areas. The difference of the energies of the wires in β-Sn phase of different orientations tends to zero with increasing of the wire size.

SiNWs in β-Sn phase display metallic properties like bulk β-Sn silicon. In the case of bulk β-Sn silicon obtained results agree with previously reported theoretical [18], [19] and experimental [13], [14] papers. It should be mentioned that metallic nature of β-Sn SiNW and α-diamond SiNW without surface passivation is different. The metallic properties of β-Sn SiNW is induced by high pressure whereas metallic behavior of α-diamond SiNW without surface passivation is induced by the surface states.

The electronic band structures of the nanowires oriented in the same direction are similar. Also the band structure of studied SiNWs is similar to parent crystal bands in the corresponding directions. For example, the band structures of bulk silicon in β-Sn phase and $\langle 001 \rangle$ silicon nanowire in β-Sn phase with cross-section area 19 Å × 19 Å = 361.2 Å$^2$ are presented in Figures 3a and 3b, respectively. Also the density of states of the studied structures is calculated. For example, DOSes of bulk silicon, of $\langle 110 \rangle$ and $\langle 001 \rangle$ silicon nanowire in β-Sn phase are presented in Fig. 3c

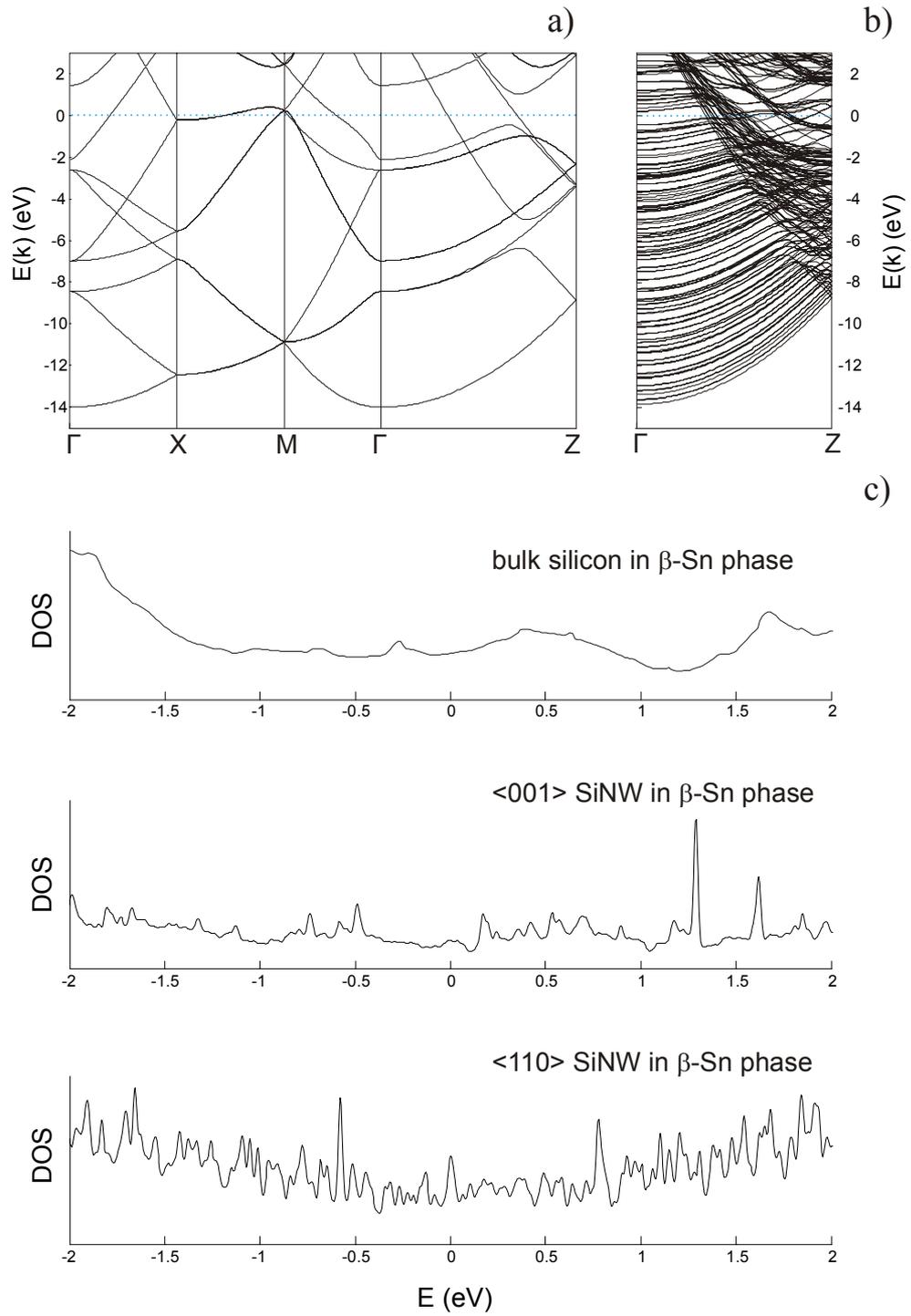

Fig. 3. The electronic band structure of a) bulk silicon in β-Sn phase, b) ⟨001⟩ SiNW in β-Sn phase. Symmetry notations of Miller and Love [28]. c) density of states of bulk silicon in β-Sn phase, ⟨001⟩ SiNW with cross-section area 361.2 Å$^2$ and ⟨110⟩ SiNWs with cross-section area 510.4 Å$^2$ in β-Sn phase. The Fermi energies are equal to zero

It was found that bulk silicon in β-Sn phase is superconductor [29] therefore we expect that β-Sn SiNW would display superconducting properties too.

This work was supported by the Russian Foundation for Basic Research (project no. 09-02-92107) and 27th RAS program. Authors are grateful to the Joint Supercomputer Center of the Russian Academy of Sciences for computer resources. The geometry of all presented structures was visualized by ChemCraft software.

## *References*

[1] M. Menon *et al.*, Phys. Rev. B **70**, 125313 (2004); I. Ponomareva *et al.*, Phys. Rev. Lett. **95**, 265502 (2005); A.K. Singh *et al.* Nano Lett. **5**, 2302 (2005); Y. Zhao and B.I. Yakobson, Phys. Rev. Lett. **91**, 035501 (2003)

[2] P.B. Sorokin *et al.*, Phys. Rev. B **77**, 235417 (2008)

[3] R.Q. Zhang *et al.*, J. Chem. Phys. **123**, 144703 (2005)

[4] J.F. Justo *et al.*, Phys. Rev. B **75**, 045303 (2007)

[5] P.V. Avramov *et al.*, Phys. Rev. B **75**, 205427 (2007)

[6] B. Marsen, K. Sattler, Phys. Rev. B **60**, 11593 (1999)

[7] R. Rurali and N. Lorente, Phys. Rev. Lett. **94**, 026805 (2005)

[8] Y. Li *et al.*, Materials today **9**, 18 (2006); J.D. Holmes *et al.*, Science **287**, 1471 (2000); Y. Huang *et al.*, Small **1**, 142 (2005);

[9] D.D.D. Ma *et al.*, Science **299**, 1874 (2003)

[10] H. K. Poswal *et al.*, Journal of Nanoscience and Nanotechnology **5**, *5*, 729 (2005)

[11] Y. Wang *et al.*, Nano Letters **8**, 2891 (2008)

[12] S. Minomura, H.G. Drickamer, J. Phys. Chem. Solids 23, 451 (1962)

[13] J.C. Jamieson, Science **139**, 762 (1963)

[14] M.C. Gupta, A.L. Ruoff, J. Appl. Phys. **51**, 1072 (1980)

[15] J.Z. Hu *et al.*, Phys. Rev. B **34**, 4679 (1986)

[16] K. Mizushima *et al.*, Phys. Rev. B **50**, 14952 (1994)

[17] R.J. Needs, R.M. Martin, Phys. Rev. B **30**, 5390 (1984)

[18] F. Zandiehnadem, W.Y. Ching, Phys. Rev. B **41**, 12162 (1990)

[19] K. Gaál-Nagy *et al.*, Comp. Mat. Science **22**, 49 (2001)

[20] P. Hohenberg and W. Kohn, Phys. Rev. **136**, 864 (1964).

[21] W. Kohn and L. J. Sham, Phys. Rev. **140**, 1133 (1965).

[22] D.M. Ceperley and B.J. Alder, Phys. Rev. Lett. **45**, 566 (1980).

[23] G. Kresse and J. Hafner, Phys. Rev. B **47**, 558 (1993).

[24] G. Kresse and J. Hafner, Phys. Rev. B **49**, 14251 (1994).

[25] G. Kresse and J. Furthmüller, Phys. Rev. B **54**, 11169 (1996).

[26] D. Vanderbilt, Phys. Rev. B **41**, 7892 (1990).

[27]   H.J. Monkhorst and J.D. Pack, Phys. Rev. B **8** 5747 (1973)

[28]   *Tables of Irreducible Representations of Space Groups and Corepresentations of Magnetic Space Groups* (Pruett, Boulder, 1967)

[29]   J. Wittig, Z. Phys. A **195**, 215 (1966)